\documentclass[graphics,floatfix,footinbib,tightenlines,nobibnotes,aps,prb,twocolumn]{revtex4-1}
\usepackage{amsmath}
\usepackage{amssymb}
\usepackage{graphicx}
\usepackage{comment}
\usepackage{soul,xcolor}
\usepackage{bbold}
\usepackage[colorlinks=true ,urlcolor=blue,urlbordercolor={0 1 1}]{hyperref}
\usepackage{tikz}
\usepackage{braket}
\usepackage[english]{babel}
\usepackage{graphicx}
\usepackage{lipsum}
\usepackage{xr}
\newcommand{\be}{\begin{equation}}
\newcommand{\ba}{\begin{align}}
\newcommand{\ee}{\end{equation}}
\newcommand{\bea}{\begin{eqnarray}}
\newcommand{\eea}{\end{eqnarray}}
\newcommand{\beq}{\begin{equation}}
\newcommand{\eeq}{\end{equation}}
\newcommand{\beqn}{\begin{eqnarray}}
\newcommand{\eeqn}{\end{eqnarray}}

\newcommand{\moire}{moir\'e }

\begin{document}

\title{Extended quantum anomalous Hall effect in moir\'e structures: phase transitions and transport}
 
\author{Adarsh S. Patri, Zhihuan Dong, and T. Senthil}

\affiliation{Department of Physics, Massachusetts Institute of Technology, Massachusetts 02139, USA}

\date{\today}

\begin{abstract}
Recent experiments on multilayer rhombohedral graphene have unearthed a number of interesting phenomena in the regime where Integer and Fractional Quantum Anomalous Hall phenomena were previously reported. Specifically at low temperature ($T$) and low applied currents, an ``Extended" Integer Quantum Anomalous Hall (EIQAH) is seen over a wide range of the phase diagram.  As the current is increased, at low $T$, the EIQAH undergoes a phase transition to a metallic state at generic fillings, and to the fractional quantum anomalous Hall (FQAH) state at the Jain fillings. Increasing temperature at the Jain fillings also leads to an evolution out of the EIQAH to the Jain state. Here we provide an interpretation of many of these observations. We describe the EIQAH as a crystalline state (either of holes doped into the $\nu = 1$ state, or an Anomalous Hall Crystal of electrons) that breaks \moire translation symmetry. At generic fillings, we show how an electric current-induced depinning transition of the crystalline order leads to peculiar non-linear current-voltage curves consistent with the experiment. At Jain fillings, we propose that the depinning transition is pre-empted by an equilibrium transition between EIQAH and Jain FQAH states. This transition occurs due to the large polarizability of the Jain FQAH states which enables them to lower their energy effectively in an applied electric field compared to the crystal states. We also discuss the finite temperature evolution in terms of the relative entropies of the crystalline and FQAH states. 
\end{abstract}

\maketitle
\section{Introduction}
Shortly after the discovery\cite{cao2018correlated,cao2018unconventional} of correlated electron physics in twisted bilayer graphene, it became clear\cite{zhang2019nearly,wu2019topological,zhang2019twisted,bultinck2020mechanism,ledwith2020fractional,repellin2020chern,abouelkomsan2020particle,wilhelm2021interplay, yu2020giant,devakul2021magic}  that many \moire materials will have nearly flat bands with Chern number making them ideal platforms to display ferromagnetism and integer/fractional quantum anomalous Hall (IQAH/FQAH) effects. The integer effect, and the associated ferromagnetism, were first observed\cite{sharpe2019emergent,chen2020tunable,serlin2020intrinsic} in a few graphene-based moire systems, as well as in some  Transition Metal Dichalcogenide (TMD) \moire systems\cite{li2021quantum,foutty2023mapping}. 
Very recently both the integer and fractional quantum anomalous Hall effect were observed\cite{cai2023signatures,zeng2023integer,park2023observation,xu2023observation} in twisted MoTe$_2$ and in pentalayer rhombohedral graphene\cite{lu2023fractional} nearly aligned with a hexagonal Boron Nitride substrate (denoted R5G/hBN). 

The standard mechanism\cite{zhang2019nearly,zhang2019twisted,bultinck2020mechanism,repellin2020ferromagnetism} for the quantum anomalous Hall phenomena in \moire materials involves the following ingredients: first the band structure is such that the bands in each valley have equal and opposite non-zero Chern number. Second, the valley (and spin, if present as an independent degree of freedom) polarizes into a ferromagnet due to interactions thereby breaking time-reversal spontaneously.  If the electron filling is such that some bands, with a net non-zero Chern number,  of the single occupied valley are fully filled, then an integer quantum anomalous Hall effect results. If the top most band is partially filled, a fractional quantum anomalous Hall effect may be seen. 

This standard mechanism accounts for the observed IQAH in many \moire materials as well as the FQAH in $t$MoTe$_2$. 
Remarkably, however, it fails in R5G/hBN which makes this system particularly interesting. 
In R5G/hBN, the IQAH is seen\cite{lu2023fractional} in the conduction band when the electrons are at a total filling $\nu = 1$. Further, the effect is seen only in a large displacement field which drives the conduction electrons away from the aligned hBN layer. In these circumstances, the non-interacting band structure of R5G/hBN is such that, even when a single valley/spin flavor is occupied, there is no band gap at filling $\nu = 1$; rather a metallic state results\cite{dong2023theory, zhou2023fractional,dong2023anomalous}. This is a result of the preferential occupation of the electrons in the layers furthest away from the aligned hBN which weakens the \moire potential.  The opening of the band gap and the emergence of a Chern number are driven by Coulomb interactions, as demonstrated in Hartree-Fock calculations of the band structure\cite{dong2023theory, zhou2023fractional,dong2023anomalous}. 
Thus, even the IQAH in R5G/hBN is the result of fascinating many-body physics. Further upon rigidly doping the Hartree-Fock band (a crude approximation) at $\nu = 1$, numerical calculations\cite{dong2023theory, zhou2023fractional,dong2023anomalous} find FQAH states at $\nu = 2/3$ and other fillings in broad agreement with experiments. Other similar $n$-layer rhombohedral graphene structures (R$n$G/hBN with $n = 4, 6, 7$) were argued\cite{dong2023theory, zhou2023fractional,dong2023anomalous} to display similar physics but at different displacement field strengths.
IQAH and FQAH states have since been seen in R4G/hBN\cite{lu2024extendedquantumanomaloushall} and R6G/hBN\cite{xie2024even}. 
 Various refinements\cite{kwan2023moir,huang2024self,huang2024fractional} of this basic theoretical picture have been explored.  

As emphasized in Ref. \onlinecite{zhou2023fractional,dong2023anomalous} (see also Ref. \onlinecite{dong2023theory}), the Hartree-Fock calculations yield a band gap with a filled Chern band even if the \moire potential is turned off completely. The resulting state breaks continuous translation symmetry, and hence is denoted an ``Anomalous Hall Crystal" (AHC). The stability of the AHC in the Hartree-Fock calculation has been explained through simplified models\cite{soejima2024anomalous,dong2024stabilityanomaloushallcrystals}. 
Whether such a state is really realized in R5G or whether a weak \moire potential is needed to enable it to occur is not currently clear\cite{dong2024stabilityanomaloushallcrystals}. For further theoretical work on anomalous Hall crystals, see Refs. \onlinecite{zeng2024sublattice,tan2024parent}. 

Very recent experimental results \cite{lu2024extendedquantumanomaloushall} on RnG/hBN show a very interesting phase diagram as a function of displacement field $D$, electron density measured through the \moire filling $\nu$, the current bias $I$,  and temperature $T$. At $\nu = 1$, the IQAH reported earlier is reproduced. However at $\frac{1}{2} < \nu < 1$, and at small $I$ and $T$, in a tilted stripe region in the $(D, \nu)$ plane, there is an ``Extended" IQAH  (EIQAH) state. A somewhat similar observation of an IQAH  was reported in Ref. \cite{waters2024interplayelectroniccrystalsinteger} but just near $\nu = \frac{2}{3}$.  For the EIQAH state of Ref. \cite{lu2024extendedquantumanomaloushall}, when $I$ is increased at low $T$, there is a transition either to a metallic state (at generic $\nu$) or to the FQAH state reported earlier (at the Jain fillings).  
Further upon increasing the temperature $T$ at low $I$ at the Jain fillings, the EIQAH phase reverts to the FQAH phase. Similar results are also found\cite{lu2024extendedquantumanomaloushall} in 4-layer rhombohedral graphene on hBN (R4G/hBN).

At generic fillings, the longitudinal differential resistance as a function of $I$ has an unusual shape: it is zero at low $I$ (as expected), and has a sharp peak at the critical current before settling to a finite value at large $I$. This shape is suggestive of some kind of depinning transition though it is seen in the resistance rather than the conductance.  Interestingly, the low $T$ critical current for the EIQAH is smaller\cite{lu2024extendedquantumanomaloushall} at the Jain fillings than at generic fillings, despite the possibility of commensuration effects strengthening any pinning processes in the former. 

The purpose of this paper is to develop an explanation of these facts based on our current theoretical understanding of RnG/hBN.   At first sight, it is tempting to interpret the EIQAH in terms of an AHC that is stabilized in a range of filling, and we will explore this below. However, a less exotic state is one where holes doped into the $\nu = 1$ state localize by forming a crystal that is pinned either by impurities, or (at commensurate fillings) by the \moire potential itself. We will show that such a hole crystal provides a natural explanation of many of the observations of Ref. \onlinecite{lu2024extendedquantumanomaloushall} (and, to a large extent, so does the AHC). 

The AHC may be thought of as a crystal formed by the electrons (counting from neutrality) with a period set by the electron density. In contrast, the hole crystal will have a period set by the hole density. We will argue that the electron and hole crystals are generically different and thus the two explanations are distinct. Thus, based on what is known so far, it is not possible to unambiguously advocate either for or against the AHC scenario for the EIQAH.  However, we present some (admittedly not very strong) arguments that suggest that the hole crystal may, in fact, be what is actually realized. 

Our work clarifies the precise sense in which the observed current-voltage characteristics at generic filling are the result of depinning phenomena. In particular we show that, due to the presence of a Hall component, the measured differential longitudinal resistance is proportional to the differential longitudinal conductance of the crystal without the Hall component. This leads to an understanding of the qualitative shape of the observed current-voltage curves. For the Jain fillings however, we propose that the depinning transition out of the EIQAH state is pre-empted by an equilibrium phase transition between crystal and FQAH states, and we present arguments for why the Jain state is favored by the current.

The remainder of the paper is organized as follows.
In Sec. \ref{sec_eiqah}, we describe the two possible classes of crystalline states that are pertinent to the EIQAH: Hole-crystal and Anomalous Hall crystal.
In Sec. \ref{sec_depinning}, we discuss the transport signatures of the depinning transition of the hole crystal (at incommensurate fillings) and the field-induced EIQAH-FQAH transition at Jain fillings of the hole-crystal.
In Sec. \ref{sec_ahc_transport_pt} we discuss the transport and phase transitions from the perspective of the anomalous Hall crystal.
In Sec. \ref{sec_finite_temp_pt}, we consider the finite temperature transitions between the EIQAH and FQAH states (at the Jain fillings) and discuss the possibility of  an explanation in terms of the higher entropy of the FQAH state.
Finally, we conclude in Sec. \ref{sec_discussions} and provide some future directions of exploration.

\section{Extended Integer quantum anomalous Hall region}
\label{sec_eiqah}


We begin with some simple general considerations on the EIQAH state at fillings $\frac{1}{2} < \nu < 1$. 
Generically, for any $\nu <1$, the partial filling of the \moire unit cell will naturally break the \moire translation symmetry unless there is either fractionalization and the associated topological order, or there are gapless charge excitations. The former happens in, eg,  the FQAH state, and the latter happens, eg,  in a Fermi liquid metal. Clearly, the  EIQAH state is not metallic, and has integer Hall conductance. While exotic gapped states without any fractional Hall conductivity are possible\cite{musser2024fractionalization}  at fractional rational filling, we will be content to explore the simpler and more likely possibility that the EIQAH state breaks \moire translation symmetry. With this assumption, there still are two possible classes of crystalline states\footnote{Related charge-ordered states at specific commensurate fillings which show an integer quantum Hall effect have been discussed in the name of `Generalized Anomalous Hall Crystals' or `Topological Charge Density Waves'; see Ref. \cite{su2024generalized} and references therein.}: 
\begin{figure}[t]
    \centering
\includegraphics[width=0.7\linewidth]{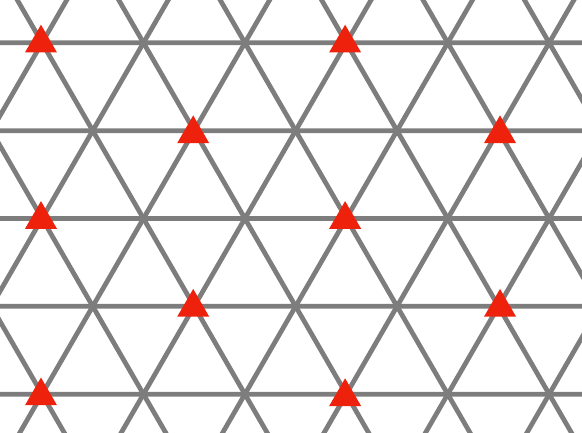}
    \caption{Hole crystal forming on top of IQAH state at rational fillings of the \moire unit cell.
    Here, a hole crystal is depicted for $\nu = 2/3$. 
    The holes shown by red triangles (of filling $\nu_h =1/3$) can form a triangular lattice, which locks to the \moire lattice.
    }
\label{fig_hole_crystal_one_third}
\end{figure}
\begin{enumerate}
    \item 
    {\bf Hole crystal } The first possibility is a (Wigner or CDW) crystal of holes forming on top of the $\nu=1$ state at a hole filling of $\nu_h = 1 - \nu$.
This hole-crystal (henceforth referred to as CDW) has its period set by $\nu_h$. At a generic $\nu_h$, this crystal will be pinned by impurities in the device. At commensurate fillings, it will lock to the \moire lattice. Note that the holes not only observe the bare \moire potential, but also the self-consistently generated electronic potential from the many-body state at $\nu = 1$. Thus the effective periodic potential seen at the \moire period will be enhanced over the bare \moire potential, and we expect the locking to be strong. An example of a commensurate hole crystal at $\nu = 2/3$ (i.e. $\nu_h = 1/3$) is shown in Fig. \ref{fig_hole_crystal_one_third}: the hole-crystal forms a triangular lattice, where the charge density resides on select sites of the \moire unit cell and gains commensuration energy from the underlying \moire potential.
 For other rational fillings $\nu = 3/5$ (i.e. $\nu_h = 2/5$) the hole-crystal may not necessarily be a triangular lattice, but instead may be determined by the Coulomb interactions on the lattice. 

\item 
{\bf Anomalous Hall crystal}
The second possibility is that of an anomalous Hall crystal (AHC)$_\rho$ which is formed by electrons at filling $\nu$.
Here the subscript denotes that the period of the electronic crystalline state is determined by the electronic density. Clearly this state also breaks \moire translation symmetry for any $\nu \neq 1$.  However the electrons forming this state see only the weak bare \moire potential. Thus, we expect it can be described essentially as the (AHC)$_\rho$ of the continuum system that is only weakly perturbed by the \moire potential. The crystal structure of (AHC)$_\rho$ will hence be triangular, and such that each unit cell of this triangular lattice has one electron. It is easy to see that the fillings at which this triangular lattice is commensurate -- and in the process gain some commensuration energy -- with the \moire lattice are $\nu = (n^2 + m^2 + nm)^{-1}$, where $n,m \in \mathbb{Z}$. \footnote{If we allow multiple electrons per CDW unit cell, then the fillings can be $\frac{p}{m^2+n^2+mn}$ with $m,n,p\in \mathbb{Z}$, which can fall into the regime of $\frac{1}{2}<\nu<1$. We note that for $p=m^2+n^2+mn-1$, this is the same as the hole crystal, where holes form a triangular lattice commensurate with \moire, with one hole per CDW unit cell. However, it is energetically unlikely that a crystal state in a continuum system has multiple electrons per unit cell.} 
This entails that the only fillings where the (AHC)$_\rho$ is commensurate with the \moire lattice are either at $\nu = 1$ or at $\nu \leq 1/3$.  Thus, in the range of interest $\frac{1}{2} < \nu < 1$, the (AHC)$_\rho$ is an incommensurate triangular lattice that `floats' on top of the \moire lattice. It will only be pinned by impurities. 

\end{enumerate}

The crystal states we are concerned with in this paper should be distinguished from the `parent' (possibly moir\'e-enabled \cite{dong2024stabilityanomaloushallcrystals}) Anomalous Hall Crystal discussed for the $\nu = 1$ state itself. This parent crystal has a period set by the \moire lattice spacing. At $\nu = 1$ this crystal period is the same as the period set by the electron density but is different at other $\nu$. 
To emphasize this, we denote this parent crystal as (AHC)$_M$. If such a parent crystal forms at $\nu \neq 1$, there will inevitably be extra holes or electrons to make up the right density. These will then form many-body states of their own. Thus the IQAH component of the hole crystal state envisaged above should  be viewed as (AHC)$_M$. Similarly, the FQAH states (at least for $\nu > 1/2$) are most usefully discussed in the hole picture, and hence thought of as a combination of (AHC)$_M$ and an FQAH state of holes. Beyond setting the stage for these many-body states to form at $\nu \neq 1$, the (AHC)$_M$ does not play a direct major role in this paper, and hence we will not discuss it further.

At $\nu \neq 1$ and low current bias, either the hole or electron crystals discussed above will stay pinned, due either to impurities or to commensuration or both,  and will have an integer quantum anomalous Hall effect. 
We identify these with the EIQAH seen in experiments. In the following, we analyze their properties in more detail focusing on the depinning phenomenon at generic filling, and their transition to FQAH states at the Jain fillings. 
 
\section{Transport and phase transitions  of the hole-crystal}
\label{sec_depinning}

In this section, we focus on the hole crystal and discuss the transition out of this state induced upon increasing the electric current. 

\subsection{Depinning transition at incommensurate fillings}

\begin{figure}[t]
    \centering
\includegraphics[width=\linewidth]{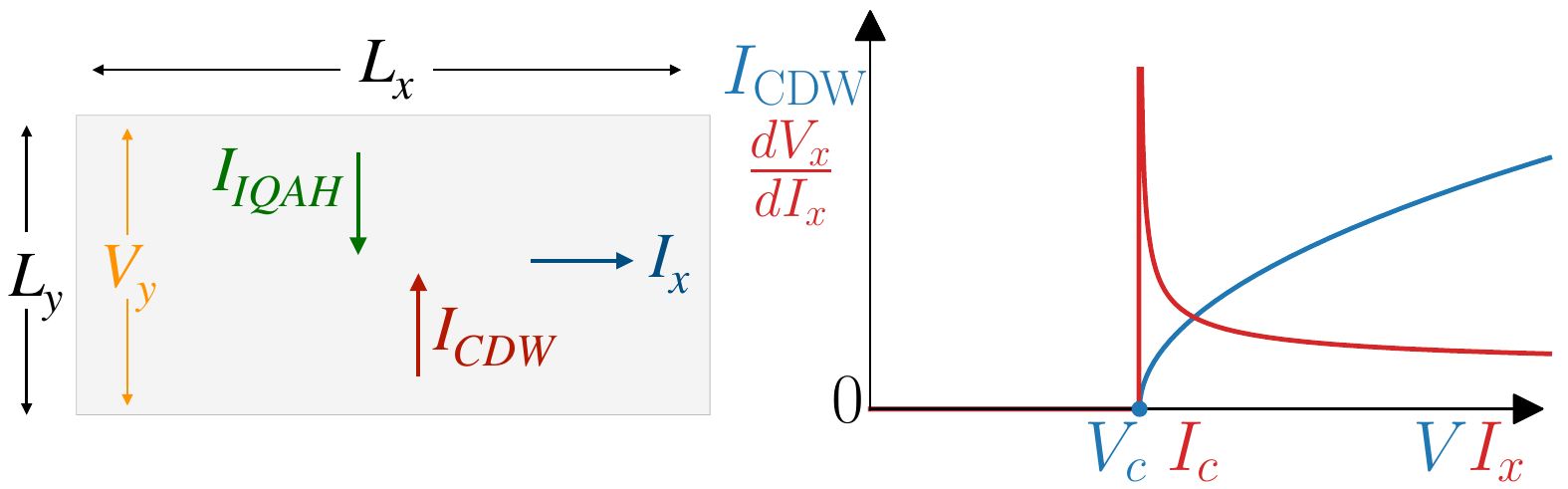}
    \caption{Depinning Transition of Hole-Crystal (CDW). Left: schematic of sample illustrating the interplay of the CDW and IQAH component at the hole-crystal depinning transition, where a counter-propagating $I_{\text{IQAH}}$ current cancels the $I_{\text{CDW}}$ sliding current. In actuality, the CDW component slides at an angle $\theta_H$ with respect to the $\hat{x}$-direction, with the generated IQAH current component in the direction perpendicular to the CDW current direction to yield a net current flowing along $I_x$.
    Right: CDW current (blue) as a function of Hall voltage and differential resistance (red) as a respective function of Hall voltage ($V$) and bias current ($I_x$) depicting the depinning transition for critical exponent $\beta = 2/3$. }
   \label{fig_depinning_transition}
\end{figure}

At incommensurate fillings, the many-body state can be imagined as being segmented into an integer quantum anomalous Hall component ($\nu = 1$) and a hole-crystal ($\nu_h$).
We consider running a current $I_x$ along the $\hat{x}$ direction in a sample of dimensions $L_x \times L_y$ as depicted in Fig. \ref{fig_depinning_transition}.
The sample geometry is such that leads are connected in the $\hat{x}$-direction with an open circuit in the $\hat{y}$-direction.
The hole-crystal, as discussed in Sec. \ref{sec_eiqah} is pinned by the disorder (at low currents). 
 Then the current is carried by the IQAH component, which leads to a Hall voltage $V_y = \frac{h}{e^2}I_x$. The corresponding electric field is felt by the hole crystal. As is well known, a pinned crystal will, when subject to a large enough electric field, start sliding. The evolution to the sliding state at $T = 0 $ happens through a sharp dynamical phase transition. We discuss below the subsequent longitudinal differential resistance.  

We first give a simplified treatment that captures the essential physics. As the electric field is along $\hat{y}$, the crystal component will slide along $\hat{y}$. (Here we are ignoring any anomalous velocity of the sliding hole crystal). Just above the threshold electric field, the sliding crystal will thus lead to a small electric current $I_{y, \text{CDW}}$ along the $\hat{y}$ direction, which must be canceled by an equal and opposite current $I_{y,\text{IQAH}} = - I_{y, \text{CDW}}$ from the IQAH component due to the open circuit along $\hat{y}$. 
In turn, $I_{y, \text{IQAH}}$ generates a voltage along $\hat{x}$,
\begin{align}
V_x (I_x) = - \frac{h}{e^2} I_{y,\text{IQAH}} = \frac{h}{e^2} I_{y,\text{CDW}}\left( V_y = \frac{h}{e^2} I_x \right),
\end{align}
where the minus sign in the first equality is due to the direction of the current flow, and in the last equality we have used the aforedescribed Hall relation.
We thus arrive at the differential (longitudinal) resistance,
\begin{align}
\frac{dV_x}{dI_x} = \left(\frac{h}{e^2}\right)^2 \frac{d I_{y,\text{CDW}} (V)}{d V} {\bigg \rvert _{V = \frac{h}{e^2}I_x }}.
\end{align}
Strikingly, the differential  resistance of the combined IQAH and hole-crystal is proportional to the differential conductance of the hole-crystal.

The depinning behavior near threshold of an incommensurate crystal is an old and much studied problem in statistical mechanics. For reviews, see Refs. \onlinecite{gruner2018density,fisher1998collective,brazovskii2004pinning,reichhardt2016depinning}. The details depend on whether the sliding is in the elastic regime or in a plastic regime (where topological defects of the ordering pattern are nucleated), whether the interactions or short-range or long-range, and the pattern of charge ordering. It is expected that for weak disorder plastic deformations of the crystal may be ignored (except perhaps at very long length scales\cite{fisher1998collective,brazovskii2004pinning,reichhardt2016depinning}).  
In that case, near the threshold, $I_{\text{CDW}}(V>V_c) \sim (V - V_c)^{\beta}$, where  $\beta < 1$\cite{}.
This can lead to a singular peak in the differential resistance (for $\beta<1$), as shown in Fig. \ref{fig_depinning_transition} in agreement with the experimental observation of the sharp transition feature at generic fillings.
The combination of the depinning of the hole-crystal in conjunction with the interplay with the IQAH component leads to this feature. 

The above treatment, though capturing the essentials of the interplay of the hole-crystal and IQAH component, ignored the sliding of the hole-crystal along the $\hat{x}$-direction by the Hall voltage $V_x$.
We  define the electric field components $E_{x,y} = V_{x,y} / L_{x,y}$, where $V_{x,y}$ are the voltage  drops along the $\hat{x},\hat{y}$-directions; and the corresponding current densities  $J_{x,y} = I_{x,y} / L_{x,y}$. 
The total current density is composed of the IQAH and CDW components, $\vec{J} = \vec{J}_{\text{CDW}} + \vec{J}_{\text{IQAH}}$ such that, ${J}_y = 0$ due to the open circuit in the $\hat{y}$-direction.
The net electric field is generated by the IQAH component and is thus perpendicular to the total $\vec{J}^{\text{IQAH}}$ i.e. $\vec{E} = \frac{h}{e^2} \hat{z} \times \vec{J}_{\text{IQAH}}$, where $\vec{J}_{\text{IQAH}}$ is the current density associated with the IQAH component. When the CDW component slides, there will be a current along the electric field direction\footnote{Recall our assumption that we ignore any anomalous velocity of the hole crystal.}: $\vec{J}_{\text{CDW}} \parallel \vec{E}$.

 Defining the Hall angle $\theta_H$ (angle between total electric field and current), and 
projecting the total current ${J}_x$ along and perpendicular to the direction of the electric field yields, 
\begin{align}
    |\vec{J}_{\text{CDW}}| &= J_{\text{CDW}} = J_x \cos(\theta_H) \\
    |\vec{J}_{\text{IQAH}}| &= J_{\text{IQAH}} = J_x \sin(\theta_H).
\end{align}
The incommensurate hole crystal will have a non-linear current-voltage relationship $J_{\text{CDW}} = f(E)$ where the function $f$ will be non-zero only beyond a sharp threshold value of its argument. Using  $|\vec{E}| = \frac{h}{e^2} J_{\text{IQAH}}$, we arrive at the self-consistent equation for the total current,
\begin{align}
    J_x \cos(\theta_H) = f \Big(\frac{h}{e^2} J_x \sin(\theta_H) \Big).
    \label{eq_hx_iqah}
\end{align}
Equation \ref{eq_hx_iqah} permits the Hall angle to be extracted from the magnitude of the total current, $\theta_H = \theta_H(J_x)$.

For small currents,   $f=0$, which leads to $\theta_H = \frac{\pi}{2}$. 
For currents just above the depinning threshold, we take $f(E) \approx A(E-E_c)^{\beta}$ , where $E_c$ is the depinning electric field, $\beta < 1$ is the critical exponent defined above, and $A$ is a constant.
Since the critical current just above the threshold is small, we can assume $\theta_H \approx \pi/2 - \delta$, where $\delta$ is taken to be small.
Expanding Eq. \ref{eq_hx_iqah} in the small $\delta$ limit yields $V_x = \tilde{A} (I_x - I_c)^{\beta}$, where $\tilde{A}$ is a constant.
We thus arrive at the same singular form of the differential longitudinal resistivity as in the simplified treatment.
Similarly, we can obtain the correction due to the sliding of the hole-crystal to the Hall resistivity of $\sim \Big[(h/e^2)J_x - x_c\Big]^{2 \beta - 1}$.

\subsection{Current-induced EIQAH-FQAH transition at Jain fillings}

At the specific Jain fillings of the \moire unit cell, the current-driven transition from the IQAH state to the FQAH state occurs at a smaller critical current than at the generic incommensurate fillings. 
This is surprising because, as described in Sec. \ref{sec_eiqah}, due to the enhanced role of the \moire potential at rational fillings,  commensuration effects, and hence the depinning thresholds,  are expected to be larger.
Thus, we propose that the current-induced transition at the Jain fillings is not a depinning of the hole crystal but rather is an equilibrium transition where the IQAH component is unchanged, while the ground state of the holes changes from the commensurate crystal to an FQAH state.

Consider the Jain FQAH state at filling $\nu = p/(2p+1)$, where for $\nu > 1/2$ this entails $p<0$.
This state can be regarded as the particle-hole conjugate of the Jain state of holes at filling $\nu_h = 1 - p/(2p+1) = (p + 1) /(2p+1)$. 
This allows us to regard the FQAH state at the Jain fillings as an IQAH + hole-Jain state.
For example, the $\nu = 2/3$ Jain state is equivalent to the IQAH + ($\nu_h = 1/3$) hole-Jain state; similarly, the $\nu = 3/5$ Jain state is equivalent to the IQAH + ($\nu_h = 2/5$) hole-Jain state.
Thus, at low bias currents, the two competing many-body states at these fillings are (a) IQAH + hole-crystal and (b) IQAH + hole-Jain state.
It is important to reiterate that the $\nu = 1$ IQAH state is the same in both many-body states.

Since the current in a quantum Hall liquid is carried by its edge state (and has no dissipation), the many-body state at a non-zero current is an equilibrium state\footnote{This does not violate the statement -- known as Bloch's second theorem -- that typically equilibrium many-body states will have zero electric current. A sophisticated discussion of a loophole to Bloch's second theorem is in Ref. \onlinecite{else2021critical}, and this loophole applies to chiral quantum Hall edge states (see also Ref. \onlinecite{kapustin2019absence}). } and we can ask about its ground state energy.  Thus our proposal is that the depinning transition at incommensurate fillings is pre-empted at the Jain fillings by the transition to the Jain FQAH states.

The transition between a hole-crystal and hole-Jain state may occur under the tuning of external knobs, such as displacement field, even at zero bias current.
This transition may even be allowed to be continuous \cite{song2024phasetransitions}, but here we will consider the more conventional possibility that it is a first-order transition.  
Let $g$ be a parameter (eg,  a displacement field) that tunes across this first-order transition at $T = I = 0$.  At the transition point, $g = g_c$, the ground state energy per hole is equal between the two states $E_{\text{CDW}} (g = g_c) = E_{\text{hole-Jain}} (g = g_c)$.
For $g<g_c$ (the evident state of the experimental sample at lowest temperatures), $E_{\text{CDW}} (g < g_c) < E_{\text{hole-Jain}} (g < g_c)$.
Nonetheless, we expect that the two energies (per hole)  to be close to each other, as is known from the competition between crystal and liquid states in standard Landau Levels, as well as other contexts (see, eg, the discussion in Ref. \onlinecite{dong2024stabilityanomaloushallcrystals} and references therein). 

\begin{figure}[t]
    \centering
\includegraphics[width=0.8\linewidth]{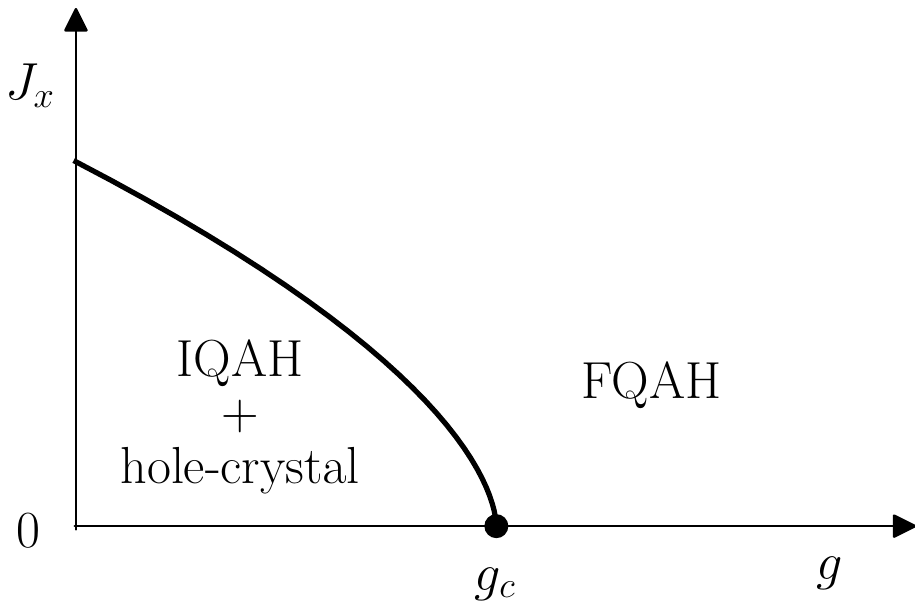}
    \caption{Schematic of the current ($J_x$) induced phase transition from IQAH + hole-crystal to FQAH state. $g$ denotes a parameter that tunes across a first-order transition at $T = I = 0$. 
    The first-order phase boundary bends towards the IQAH + hole-crystal, indicative of the enhanced dielectric susceptibility of the Jain states.}
\label{fig_iqah_fqah_pt}
\end{figure}

We now consider the relative stability of the hole crystal and the hole-Jain states in the presence of an electric current   $\vec{J} = J_x \hat{x}$. The induced Hall electric field will lead to an electric polarization which will lower the energy of either state, and will affect their relative stability,  as we discuss below. 
We provide a schematic phase diagram of the current-induced phase transition from IQAH + hole-crystal to FQAH state in Fig. \ref{fig_iqah_fqah_pt}.

\subsubsection{IQAH + hole-crystal state}

The IQAH background component is the sole carrier of the current which gives rise to a perpendicular Hall electric field $E_y = \frac{h}{e^2}J_x$.
Since the applied bias current is smaller than the depinning threshold current, the hole-crystal does not slide but instead develops an electric polarization.
The modification of the ground state energy due to this polarization will (for small currents) be, 
\begin{align}
    E_{\text{CDW}}(J_x) = E_{\text{CDW}}(J_x=0) - \frac{\chi_{\text{CDW}}}{2} \left(\frac{h}{e} \right)^2 J_x ^2,
\end{align}
where $\chi_{\text{CDW}}$ is the dielectric susceptibility of the hole-crystal. We expect $\chi_{\text{CDW}}$ to be a smooth function of the filling $\nu$.

\subsubsection{IQAH + hole-Jain state}

For the IQAH + hole-Jain state, the current (and consequently conductivity) is composed of the contribution from both IQAH and the hole-Jain components,
\begin{align}
    \sigma_{xy} &= \sigma_{xy}^{\text{IQAH}} + \sigma_{xy}^{\text{hole-Jain}} \nonumber \\
    & = \frac{e^2}{h} - \frac{e^2} {h} \frac{p+1}{2p+1} \nonumber \\
    & = \frac{e^2}{h} \frac{p}{2p+1}
\end{align}
where the negative sign in the second line is due to the hole charge.
We thus obtain the corresponding electric field, $E_y = \frac{h}{e^2} \frac{2p+1}{p}J_x$.
The hole Jain state will also develop an electric polarization due to this Hall electric field,
\begin{align}
    E_{\text{hole-Jain}}(J_x) &= E_{\text{hole-Jain}}(J_x=0) \nonumber \\
    & - \frac{\chi_{\text{hole-Jain}}}{2} \left(\frac{h}{e} \right)^2 \left(  \frac{2p+1}{p} \right)^2 J_x^2,
\end{align}
where $\chi_{\text{hole-Jain}}$ is the corresponding dielectric susceptibility.
For large $|p|$ (i.e. fillings closer to $\nu = 1/2$) the electric field responsible for the polarization of the hole-Jain state is two times larger than the electric field polarizing the hole-crystal due to the reduced Hall conductivity. 

Unlike the hole-crystal's dielectric susceptibility, the dielectric susceptibility of the Jain state will grow with increasing $|p|$, and eventually diverge.  Thus the hole-Jain state for large $|p|$ can lower its energy in the presence of a current more effectively than the hole-crystal can. It then follows that for $g < g_c$, there will be a transition at a non-zero current where the ground state flips from the EIQAH to FQAH states. 
 To understand the Jain state's susceptibility, first recall that the termination of the Jain series as $|p| \rightarrow \infty$ is the $\nu = 1/2$ compressible composite Fermi liquid (CFL) where the static wavevector dependent dielectric function $\chi(q) = \frac{\kappa}{q^2}$ for small $q$ (here  $\kappa$ is the electronic compressibility). If the electron density changes by $\delta \rho$ to move away from $\nu = 1/2$, the composite fermions see an effective magnetic field $B^* = -4\pi \delta \rho$. There is a corresponding length scale ${l_{B^{*}}}^2 = \frac{1}{B^*}$.  The  small $q$ divergence of $\chi(q)$ will be cut-off at a momentum scale $\frac{1}{{l_{B^*}}^2}  = {B^*}$ leading to a dielectric susceptibility 
\begin{align}
    \chi_{\text{hole-Jain}} = \kappa  l_{B^*}^2   \sim |p|, 
\end{align}
where we have employed the fact that $|B^*| = B/(2p+1)$ at the Jain fillings. 
Thus, as promised,  for Jain fillings close to $\nu = 1/2$ (i.e. large $|p|$) the dielectric susceptibility is very large.
This ensures that the hole-Jain state will always be stabilized by a small $J_x$ over the hole-crystal for Jain fillings sufficiently close to $\nu = 1/2$.
A detailed energy comparison away from this asymptotic limit ultimately requires an accurate model for the hole-crystal and hole-Jain state, and is beyond the purview of our study.
We demonstrate in Appendix \ref{app_sec_dielectric_susceptibility_cfl} a mean-field composite fermion treatment for the energy decrease of the Jain state, which confirms the above-presented argument.
We emphasize that the above argument for the dielectric susceptibility is beyond the mean-field treatment.

\section{Transport and phase transitions of the Anomalous Hall Crystal} 
\label{sec_ahc_transport_pt}

Next we turn to the (AHC)$_\rho$ state which is an electron crystal (with a period determined by the charge density), and whose occupied bands have non-zero Chern number $C = 1$. At low current strengths $\vec J = J_x \hat{x}$, the (AHC)$_\rho$ will stay pinned by impurities. The non-zero Chern number leads to a Hall electric field $E_y = \frac{h}{e^2} J_x$. With increasing current, this electric field will also increase and, like in any other pinned crystal, will eventually lead to a sharp depinning transition at $T = 0$. 
Beyond the threshold, the AHC will slide but now in a direction that is different from that of the electric field. This is because the sliding AHC will have an anomalous velocity (just like it does in the pinned state). We analyze the transport properties across the depinning transition below, and show that it will have the same qualitative behavior as the incommensurate hole crystal (and the experiments).
Subsequently, we will discuss the situation at the Jain fillings. 

\subsubsection{Depinning transition of  Anomalous Hall crystal $\text{(AHC)}_{\rho}$}

Consider the current response of the (AHC)$_\rho$  to an electric field  $\vec E$. In general we can write this as
\begin{equation}
    \vec J = \sigma_{xx}(E) \vec {E} + \sigma_{xy} (E) \vec z \times \vec E .
\end{equation}
Here $\sigma_{xx}, \sigma_{xy}$ are the (non-linear) longitudinal and Hall conductivities ({\it not} the differential conductivities) that are functions of $E = |\vec E|$. For $E$ below some threshold $E_T$, we have 
\begin{equation}
    \sigma_{xx}(E < E_T) = 0; ~~\sigma_{xy} (E) = \frac{e^2}{h}.
\end{equation}
What happens beyond threshold? We expect that the longitudinal current is similar to an ordinary sliding crystal. We therefore focus on the anomalous ({\it i.e.} Hall) current. 
To understand this,  consider a sliding AHC in the complete absence of any impurity pinning which is appropriate deep into the sliding phase. In this limit, we can use a semi-classical framework for the many-particle system to obtain an effective equation of motion for the center-of-mass motion. This is detailed in Appendix \ref{app_anomalous_velocity}. We find an anomalous center-of-mass velocity,
\begin{align}
    \langle \vec{v}_{\text{CM}} \rangle = \vec{E} \times \hat{z} \bigg \langle \int \frac{d^{2}\vec{p}}{(2 \pi)^2} \mathcal{B}_{\alpha}(\vec{p}) c^{\dag}_{\alpha \vec{p}} c_{\alpha \vec{p}}\bigg \rangle,
\end{align}
where $\mathcal{B}_{\alpha}(\vec{p})$ is the microscopic Berry curvature of a microscopic electron of flavor $\alpha$ with corresponding fermionic operators $c$. The average is evaluated in the many-body state (for vanishing electric field, $\vec{E} = 0$).
We computed this averaged Berry curvature within Hartree-Fock theory for R5G/hBN; it lies between $\approx 1 - 50$ for displacement field energies $u_d$ ranging from 10 meV to 50 meV at a range of twist angles $\theta \in [0^{\circ}, 1.5^{\circ}]$. The center-of-mass anomalous velocity of the AHC determines the Hall current $\vec J_{\text{Hall}} = \rho \langle \vec{v}_{\text{CM}} \rangle$ where $\rho$ is the charge density.  The crucial point is that $\sigma_{xy} (E \gg E_T) = \rho \neq 0$ but is also not quantized. Across the continuous dynamical transition at $E_T$, we expect that $\sigma_{xy}(E)$ will evolve continuously from its quantized value for $E < E_T$ to its unquantized value for $E \gg E_T$. We will however not attempt to describe the details of the critical behavior of the anomalous velocity near the threshold. Once we accept that $\sigma_{xy}(E \approx E_T)$ is close (or exactly equal) to $e^2/h$, the analysis of the longitudinal differential resistivity becomes identical to that in Sec. \ref{sec_depinning} for the incommensurate hole crystal.  Thus there will be a peak in the longitudinal differential resistance at low $E$, and pass through a peak at the threshold, before settling to a finite non-zero value in the sliding phase.

Thus we see that the (AHC)$_{\rho}$ possesses both a longitudinal conductivity qualitatively similar to the incommensurate hole-crystal.  The distinction with the IQAH + hole-crystal picture is that here the (AHC)$_{\rho}$ contains both longitudinal and Hall components in one electronic entity. 

The detailed behavior of the Hall component near the threshold is an interesting theoretical problem for the future.  

\subsubsection{Current-induced transition to Jain FQAH }
The discussion on the current-induced transition from the hole crystal to the Jain FQAH can be taken over without any modification to the (AHC)$_\rho$ as well, and we will mainly highlight some differences.  In short, we propose that the depinning transition of the (AHC)$_\rho$ is pre-empted by a current-induced first-order transition to the  Jain FQAH. In the presence of a current $J_x$, the pinned (AHC)$_\rho$ will experience the Hall electric field and will develop a polarization that lowers its energy of $\mathcal{O}(J_x^2)$. The coefficient is proportional to the dielectric susceptibility which will be a smooth non-diverging function of $\nu$. Thus the large susceptibility of the Jain FQAH state can help stabilize it over the (AHC)$_\rho$ as detailed in the discussion on the hole crystal. 

However, we find this proposal somewhat less compelling in the (AHC)$_\rho$ than in the hole crystal. As we emphasize below, in the latter, the effect of the \moire potential is expected to be enhanced, and hence the depinning thresholds at the Jain fillings may be large enough that they can be pre-empted by the transition to the FQAH state. In contrast, in the (AHC)$_\rho$, there is nothing special about the Jain fillings as far as pinning thresholds go. They are expected to evolve smoothly across these fillings. It is then not clear that the depinning transition can be pre-empted in the manner described in the previous paragraph. 
A second concern is that while the hole-crystal and the Jain state can reasonably be expected to have a close competition ({\it i.e.} have close energies/particle) even away from the first order transition point, this is less clear for the (AHC)$_\rho$ and the Jain state. These two states are rather different from each other, and any closeness of their energies may be specific to a narrow region near their presumed first-order transition. 

\section{Finite temperature transitions between EIQAH and FQAH at the Jain fillings}
\label{sec_finite_temp_pt}

The EIQAH is stabilized at the base temperature\cite{lu2024extendedquantumanomaloushall} ($\sim$10mK). As temperature rises to $T_c \sim100$mK, a transition from EIQAH to FQAH is observed for the Jain fillings $\nu=\frac{2}{3},\frac{3}{5}$.  It is possible that this phenomenon has a mundane explanation due to experimental artifacts such as equilibration issues between the contacts and the edge modes at low $T$ \cite{young_private_comm}. Here however we will explore the possibility of an explanation where such a transition is an intrinsic bulk phenomenon. Then the experimental observation  suggests that the FQAH state is slightly higher in energy than the EIQAH state, but it also has higher entropy, which at higher temperatures reduces its free energy.

Certainly, this entropy difference should be attributed to low-energy excitations. 
The pertinent excitations are those at an energy scale lower or at least similar to the transition temperature $T_c \sim 100$mK$\sim0.01$meV, so that they contribute significant entropy at $T_c$. Candidate excitations are charged quasiparticles or neutral collective modes. The charged gap of the hole crystal and the (AHC)$_{\rho}$ state should be of order the Coulomb repulsion. For the FQAH, at least close to $\nu = 1/2$, the charge gap will become small (order $\frac{1}{|p|}$ in composite fermion mean-field theory). Away from this limit, the gaps are not known accurately but we can take guidance from existing calculations. According to numerics, within the approximation that the interaction-induced Chern band is treated as frozen, the FQAH at $\nu = \frac{2}{3}, \frac{3}{5}$ has a gap of $\sim 1$meV\cite{dong2023theory,dong2023anomalous,zhou2023fractional}.  The most important neutral modes are the phonons of the broken translation symmetry. At Jain fillings, there are three competing states: (1) electron crystal (AHC)$_\rho$, (2) IQAH + hole-crystal, and (3) IQAH + hole-Jain state.  In (2) and (3), the IQAH component really is (AHC)$_M$. 
In the absence of a \moire potential, all of these break continuous translation symmetry and therefore have gapless phonons.  
In the presence of a \moire potential, a phonon gap is opened of size determined by the effective \moire potential strength. Note in particular the enhancement of the \moire potential by the underlying IQAH state for (2).  However, the phonon for (1) remains gapless since the (AHC)$_\rho$ lattice is incommensurate with the \moire lattice.

Armed with this understanding, we can discuss the finite-$T$ evolution of the EIQAH phase. Due its gapless phonon, the (AHC)$_\rho$ will have a high entropy at low-$T$ compared to the other two states. Thus if it is already the preferred ground state, it is unlikely to yield to the FQAH on the scale of $100$ mK.  
However, in the experiment, the EIQAH transitions into FQAH at higher temperatures.
This suggests that the underlying low-temperature EIQAH state is more likely to be the IQAH + hole-crystal, as it does not manifestly overwhelm the FQAH state at higher temperatures, unlike the (AHC)$_\rho$. The high temperature stabilization of the FQAH relative to the hole-crystal is easy to understand for $\nu$ close to $1/2$ where the charged excitations of the former have increasingly small gaps. 
Well away from $\nu = 1/2$, based on the estimates above, the charged excitations are unlikely to play a role at the $10$ mK scale. Evaluating the relative stability of the hole crystal and FQAH states at non-zero $T$ will have to await a better understanding of the excitation spectrum of either state.

\section{Discussion}
\label{sec_discussions}

In this work, we examined the nature of the extended integer quantum anomalous Hall (EIQAH) phase\cite{lu2024extendedquantumanomaloushall,waters2024interplayelectroniccrystalsinteger} realized at low bias currents and temperatures in rhombohedral graphene \moire structures. For very general theoretical reasons, the EIQAH state away from $\nu = 1$  breaks \moire translation symmetry. Thus the main question is the nature of the resulting crystalline state, its depinning transition, and its competition with the FQAH state at Jain fillings. We explored two possible crystalline states away from $\nu = 1$: 
  (a) a crystal of holes doped into the $\nu = 1$ state  (IQAH + hole-crystal), and (b) an electron crystal with a quantized anomalous Hall effect (in its pinned state), known as the Anomalous Hall Crystal  (AHC)$_\rho$. 
The hole-crystal experiences an enhanced  \moire potential (due to the self-consistently generated electronic potential from the $\nu = 1$ background) which enhances commensuration effects. In contrast the natural (AHC)$_\rho$ state (with one electron per unit cell) is incommensurate for any filling $1/2 < \nu < 1$ where the EIQAH is seen. Thus the \moire lattice is expected to play a more important role in the hole-crystal compared to the (AHC)$_\rho$ crystal. 

At generic fillings, either crystal is pinned by impurities. The current-induced transition out of the EIQAH observed at generic fillings in the experiments\cite{lu2024extendedquantumanomaloushall} can be understood as a depinning of the crystal.  For the IQAH + hole-crystal,  the interplay of the sliding hole-crystal and the IQAH component leads to a sharp peak in the differential resistance at threshold. In the  (AHC)$_\rho$ picture, the threshold behavior can similarly be understood with the only difference being that the Hall and longitudinal components are due to the same electrons. 

At the Jain fillings, we proposed that the current-induced transition out of the EIQAH is not a depinning transition of the crystal. Rather the depinning transition is pre-empted by a ground state phase transition to the Jain FQAH state. For the hole-crystal, it is natural to expect that it competes closely with the FQAH state (which can be viewed as a hole Jain state) at the same filling. Thus even if the hole-crystal wins at zero current, the energy balance can be tipped in an electric field depending on which of these states has a higher dielectric susceptibility. We analyzed this effect and showed that for fillings close to $\nu = 1/2$, the Jain state will have a strongly enhanced susceptibility (inherited from the proximity to the compressible composite Fermi liquid at $\nu = 1/2$). The hole-crystal does not have any such enhancement, and hence, an external current (and the associated Hall electric field) can drive a transition from the hole-crystal to the Jain FQAH state. In principle, a similar picture may also apply to the competition between (AHC)$_\rho$ and the FQAH state. However, it is less clear if these states are close competitors. 
 
Finally, we discussed the finite temperature transition to the FQAH state in the framework of entropy gain of the FQAH states relative to the EIQAH state. We suggested that the pertinent excitations for the entropy may be phonon modes of the various crystals as they are likely to have lower energy than the charged excitations. Unlike the hole crystal and the Jain FQAH, the (AHC)$_\rho$, being incommensurate with moir\'e,  will have an ungapped phonon spectrum even at the Jain fillings, and hence have a higher entropy than the FQAH state. This makes it harder to understand the stabilization of the FQAH state at higher temperatures if the (AHC)$_\rho$ is realized in the EIQAH. The relative thermal stability of the IQAH hole crystal and the FQAH is hard to evaluate without a more detailed understanding of the excitation spectrum of either state than is currently available. Thus we do not have a good explanation of the finite-$T$ evolution at present, and this is a target for future work. 

We have only attempted to provide a `broad brush' discussion of the basic physics of these crystalline states and their transitions, and we leave open many detailed questions for the future. First, it will be important to take a more microscopic approach in discussing the phase competition between crystal and FQAH states at the Jain fillings. Second, there are interesting questions associated with the depinning transition in the presence of the Hall component including a detailed picture of the threshold behavior of the differential Hall conductance. Finally, our discussion of the entropy of the various competing states highlights the need for accurate microscopic studies of the excitations of these various competing states. 

\textit{Note added} -- Just before submission, we learned of Ref. \onlinecite{sarma2024thermalcrossovercherninsulator}. In this paper, they examine the thermal evolution from EIQAH to FQAH, and argue that the disorder plays an important role.

\acknowledgements

We thank Long Ju, Tonghang Han, and Zhengguang Lu for sharing and discussing their experimental data. We also thank Patrick Lee, Mehran Kardar, Cristina Marchetti, Matt Yankowitz, Andrea Young, and Ya-Hui Zhang for pertinent discussions.  TS was supported by NSF grant DMR-2206305, and partially through a Simons Investigator Award from the Simons Foundation. This work was also partly supported by the Simons Collaboration on Ultra-Quantum Matter, which is a grant from the Simons Foundation (Grant No. 651446, T.S.). 

\appendix

\section{Dielectric susceptibility within mean-field composite fermion theory}
\label{app_sec_dielectric_susceptibility_cfl}

In this section, we employ the composite fermion mean-field framework to calculate the dielectric susceptibility.
We examine the Jain state at filling $\nu_h = \frac{p+1}{2p+1}$ as being constructed from 
occupying Landau levels with composite fermions.
The effective low-energy Lagrangian is,
\begin{align}
    \mathcal{L} = \frac{1}{2 \pi} A \wedge db - \frac{2}{4 \pi} b \wedge db - \frac{1}{2\pi} b \wedge da + \mathcal{L}[\psi_{\text{cf}}, a],
    \label{eq_lagrangian_cfl}
\end{align}
where $A$ is the external background gauge field, $a,b$ are the internal gauge fields, and $\mathcal{L}[\psi_{\text{cf}}, a]$ is the Lagrangian involving the composite fermions and its respective gauge field $a$.
The physical electric current carried by the holes along the $\hat{x}$ direction is,
\begin{align}
    J_x ^{\text{hole}}&= \frac{1}{2 \pi} e_y^b = \frac{1}{2\pi} \left( -\partial_y b_0 - \partial_0 b_y \right).
    \end{align}
The total current $J_x = J_x^{\text{hole}} + J_x^{\text{IQAH}}$ can be related to the current carried by the holes via the background electric field $E_y$,
\begin{align}
    J_x^{\text{hole}} = \frac{p+1}{p}J_x \implies \frac{e_y^b}{E_y} = \frac{p+1}{2p + 1}.
\end{align}
Varying the Lagrangian Eq. \ref{eq_lagrangian_cfl} with respect to $b$ in the presence of the external magnetic field along the $\hat{y}$-direction yields,
\begin{align}
    e_y^a &= E_y - 2 e_y^b \\
    &=- \frac{E_y}{2p+1} 
\label{eq_eyb_ey_relation}
\end{align}
where we have used that $dA = -E_y$ and similarly for the internal gauge fields, and in the second equality we employed Eq. \ref{eq_eyb_ey_relation}.

The composite fermions residing in experience the electric field $e_y^a$.
Within first quantization, the coupling of the electric field to the composite fermion location is, $H_e = - y e_y^a$.
Within Landau levels, $\langle H_e \rangle = 0$, indicating the lack of dipole moment in the ground states.
The dielectric susceptibility is computed with second-order perturbation theory,
\begin{align}
    \frac{\Delta \epsilon}{N} = \sum_{m>n} (e_y^a)^2 \frac{| \langle m | y | n \rangle |^2}{- \hbar \omega_c^* (m-n)},
\end{align}
where $\omega_c^*$ is the cyclotron frequency associated with the effective magnetic field $B^*$ experienced by the composite fermions.
For Landau levels, only for $m = n+1$ is the matrix element non-vanishing (recall $m>n$ in the summation), and so
\begin{align}
   \frac{\Delta \epsilon}{N} &= - (e_y^a)^2 \frac{m^*}{(B^*)^2} c \\
   & = - E_y^2 |p| m^* c,
\end{align}
where $c$ is a numerical factor, and we have once again employed Eq. \ref{eq_eyb_ey_relation} for $e_y^a$.
It follows that the dielectric susceptibility is,
\begin{align}
    \chi_{\text{hole-Jain}} \sim |p| m^*
\end{align}
Assuming the effective mass stays constant as $|p| \rightarrow \infty$, then dielectric susceptibility diverges as $\nu \rightarrow 1/2$.
Thus the Jain state will always be stabilized by a small current $J_x$ for sufficiently large $|p|$ i.e. for fillings sufficiently close to $1/2$.

\section{Anomalous drift velocity within semi-classical framework}
\label{app_anomalous_velocity}

In this section, we derive the anomalous drift velocity of the (AHC)$_{\rho}$ within a semi-classical framework. Consider the microscopic semi-classical equations of motion for an electron of the continuum rhombohedral-stacked $n$-layer graphene (R$n$G) Hamiltonian,
\begin{align}
    H = \sum_i \epsilon(\vec{p}_i) + \frac{1}{2} \sum_{i \neq j} V(\vec{r}_i - \vec{r}_j) - \vec{E} \cdot \sum_i \vec{r}_i,
\end{align}
where $\epsilon(\vec{p}_i)$ is the dispersion of the R$n$G electron, $V(\vec{r}_i - \vec{r}_j)$ is the interaction between the electrons.
The final term is the microscopic electric field coupling to the electron at location $i$ in real space.
Due to the non-trivial Berry curvature of the microscopic model, the spatial coordinates $\vec{r}_i$ are the gauge-invariant kinematic coordinates, $\vec{r}_i = \vec{x}_i + \vec{\mathcal{A}}(\vec{p}_i)$.
This Hamiltonian is written in first-quantized notation.

The subsequent semi-classical equations of motion are,
\begin{align}
    \dot{\vec{p}}_i = \vec{E} - \sum_{j\neq i} \vec{\nabla}_i V(\vec{r}_i - \vec{r}_j) \\
    \dot{\vec{r}}_i = \frac{\partial \epsilon}{\partial \vec{p}_i} + \dot{\vec{p}}_i \times \Big[{\mathcal{B}}(\vec{p}_i) \hat{z} \Big]  
\end{align}
where $\mathcal{B}$ is the Berry curvaure of the microscopic R$n$G continuum model.

The center-of-mass momentum $\vec{p}_{\text{CM}} = \sum_i \vec{p}_i$ satisfies the expected equation of motion $ \dot{\vec{p}}_{\text{CM}} = N \vec{E}$, where $N$ is the total number of electrons.

The anomalous velocity contribution to the total velocity, $\vec{v}_{\text{CM}} ^{ \ a} = \frac{1}{N} \sum_i \dot{\vec{r}}_i$ is the Berry-curvature dependent term,
\begin{align}
\vec{v}_{\text{CM}}^{ \ a} =  \frac{1}{N}\sum_i \vec{E} \times \hat{z} \mathcal{B}(\vec{p}_i).
\end{align}
This operator equation is averaged over the many-body wavefunction (at $\vec{E} = 0$ for linear response),
\begin{align}
\langle \vec{v}_{\text{CM}}^{ \ a} \rangle &= \vec{E} \times \hat{z} \bigg \langle \frac{1}{N} \sum_i \mathcal{B}(\vec{p}_i) \bigg \rangle    \\
& = \vec{E} \times \hat{z}  \bigg \langle \int \frac{d^2 \vec{p} }{(2 \pi)^2} \mathcal{B}_{\alpha} (\vec{p}) c^{\dag}_{\alpha} (\vec{p}) c_{\alpha} (\vec{p}) \bigg \rangle
\end{align}
where in the second equality we have reverted to second quantization in the non-interacting band-basis in the \moire Brillouin zone, and $\alpha$ labels the bands of interest.
This expectation value is computed within the Hartree-Fock theory, where one transforms the non-interacting fermionic operators into the Hartree-Fock basis.

\bibliographystyle{apsrev4-1}
\bibliography{FQAH}

\end{document}